\documentstyle[12pt,fleqn,cite]{article}
\newcommand{\sect}[1]{\setcounter{equation}{0}\section{#1}}

\def\sxn#1{\sect{#1}}

\def\be{\begin{equation}}
\def\ee{\end{equation}}
\def\bea{\begin{eqnarray}}
\def\eea{\end{eqnarray}}

\begin{document}
\begin{titlepage}
\begin{center}
\hfill hep-th/9711048\\
\hfill CERN-TH/97-303\\
\hfill LPTENS-97/49\\
\hfill UCLA/97/TEP/25
\vskip .2in

{\Large \bf $N=(4,2)$ Chiral Supergravity in Six Dimensions and
Solvable
Lie Algebras}
\vskip .5in

\centerline{\bf Riccardo D'Auria$^{*}$,}
\vskip .2cm
{\em  Dipartimento di Fisica, Politecnico di Torino,\\ C.so degli
Abruzzi, 24, I-10129 Torino, Italy}.
\vskip .5cm
{\bf Sergio Ferrara$^{*}$ and Costas Kounnas\footnote{On
leave from Ecole Normale
Sup\'erieure, 24 rue Lhomond, F-75231, Paris Cedex 05,
France.\\e-mail address:
kounnas@nxth04.cern.ch}
\\
\vskip .2cm
{\em  Theory Division, CERN,\\CH-1211, Geneva 23, switzerland.}
\vskip .5cm
{\em * Department of Physics and Astronomy,\\University of
California, Los Angeles, USA.}}
\vskip .7cm
\end{center}

\begin{center} {\bf ABSTRACT}
\end{center}
\begin{quotation}\noindent
\baselineskip 10pt
 Decomposition of the solvable Lie algebras of maximal supergravities
in $D~$= 4, 5 and 6 indicates, at least at the geometrical level, the
existence of an $N=(4,2)$ chiral supergravity theory in $D =6$
dimensions. This theory, with 24 supercharges, reduces to the known
$N=6$ supergravity after a toroidal compactification to $D=5$ and
$D=4$. Evidence for this theory was given long ago by B. Julia. We
show that this theory suffers from a gravitational anomaly equal to
4/7 of
the pure $N=(4,0)$ supergravity anomaly. However, unlike the latter,
the absence of $N=(4,2)$ matter to cancel the anomaly presumably
makes this theory inconsistent. We discuss the obstruction in
defining  this theory in $D=6$, starting from an $N=6$
five-dimensional
string model in the decompactification limit. The set of massless
states necessary for the anomaly cancellation appears in this limit;
as a result the $N=(4,2)$ supergravity in
$D=6$ is extended to $N=(4,4)$ maximal supergravity theory.
\end{quotation}
\vskip .2in
CERN-TH/97-303\\
November 1997\\
\end{titlepage}
\vfill
\eject

\sxn{Introduction}
 Supersymmetry algebras in any dimension and their representations
are closely related to properties of spinors. Spinors of $O(D-1,1)$
are
real for $D=2,3,9$ mod 8, pseudoreal for $D=5,6,7$ mod 8, and complex
in $D=4,8$ mod 8. The above reality properties refer to spinors of a
given chirality when the dimension is even. The corresponding
supersymmetry algebras have automorphism groups $H$ (sometimes called
R-symmetry), which fall in three categories \cite{str}:
$$
i)~~O(N), ~~~~ii) ~~USp(N) ~~~{\rm and}~~~ iii)~~U(N),
$$
$N$ being the number of spinorial charges in the appropriate
dimension. We note that in $D=2$ and 6 (mod 8) algebras with
different numbers of chiral charges $N_L,N_R$ exist with
$O(N_L)\times
O(N_R)$ and $USp(N_L)\times USp(N_R)$ as $H$-groups respectively. If
one considers compactifications of higher-dimensional theories on
smooth manifolds the number of supersymmetry charges $N$ in lower
dimensions is always $2^n$ for some $n$. This leaves aside certain
theories, which can be constructed in lower dimensions, where the
number of supercharge components is \cite{ss},\cite{fk}:

$\bullet$ In $D=4$: 12, 20 and 24, corresponding to $N=$ 3, 5 and 6
supergravity with $H=U(3),~U(5)$ and $U(6)$ respectively.

$\bullet$ In $D=5$ and $D=6$ : 24, corresponding to $N$= 6 with
$H=USp(6)$ and $N=(4,2)$ with $H=USp(4)\times USp(2)$
respectively.\footnote{Instead of the notation, $N=1$ for the minimal
and $N=2$
for the maximal supersymmetry in $D=6$, we are using the chiral
notation $N=(2,0)$ for the minimal and $N=(4,4)$ for the maximal
supersymmetry;
this notation is more conveniant in $D=6$ since the  chiral spinors
in $D=6$
are Majorana symplectic\cite{ss}.}
Theories in $D=4$ and 5, have been shown to exist as consistent type
II string compactifications on asymmetric orbifolds \cite{fk},
\cite{ff}. The $N=6$ supergravity is special because it is the only
one that can be lifted to six dimensions where it becomes chiral
\cite{jul}. In $D=4$ the $N=6$ supergravity is also special when is
defined in anti de Sitter space. In fact $Osp(6\vert 4)$ gauge
supergravity
is the only one with a zero-center representation.\cite{flato}

In section 2 of this work we explore some properties of $N=(4,2)$
supermultiplets in $D=6$ by means of the chain decomposition
$N=(4,4)\rightarrow N=(4,2)\rightarrow N=(0,2)$. This allow a
derivation of
the structure of the $N=(4,2)$ supergravity, and in particular a
relation of  its scalar manifold to the decomposition of the maximal
supergravity into $N=2$ sub-theories. We show that this theory
suffers from a gravitational anomaly, which is $4/7$ of the $N=(4,0)$
pure supergravity anomaly.

In section 3 we derive the $D=5$, $N=(4,2)$ string ground state and
show why the four- and five- dimensional descendants of the anomalous
six-dimensional theory are not affected by this anomaly when one
includes massive degrees of freedom dictated by string theory.

In section 4 we analyse the two decompactification limits
$R\rightarrow \infty$, $R\rightarrow 0$ and we show the restoration
of $N=(4,4)$ in $D=6$. Our conclusions are in section 5.

\sxn{$N=2$ and $N=6$ decomposition of maximal supergravity in
$D=4$, 5 and 6 dimensions.}

Let us first consider the general decomposition of maximal
supergravities in terms of $N=2$
multiplets\cite{gst},\cite{adaf},\cite{adafft2}.
The important fact in
this decomposition is that the spin 3/2 multiplet does not contain
scalars. If we associate to the scalar manifold $\cal M$=$G/H$ a
solvable group with a Lie algebra $\cal S$ \cite{adafft1}, then we
find that the $\cal S$ vector space decomposes into $\cal S_V$+$\cal
S_H$, where $\cal S_V$ and $\cal S_H$ are solvable Lie algebras
associated to $J_{max}$=1 and $J_{max}={1\over 2}$ respectively. The
$J_{max}={1\over 2}$ \cite{adaf},\cite{adafft2},
multiplets are
hypermultiplets and  $\cal S_H$ is thus related to a quaternionic
space with holonomy $SU(2)\times H$; $H$ is the automorphism group of
the supersymmetry algebra in the corresponding dimension.

Note that in $D=6$, since the vector multiplet has no scalars, the
$J_{max}=1$ multiplet is a tensor multiplet. Our starting point is to
decompose the maximal $D=6$ supermultiplet into $N=2$ ones. This
amounts to decomposing the $N=(4,4)$ representation into $N=(0,2)$
massless representations.

The $N=(0,2)$ representations are:
\vskip .5cm
$\bullet$ {\bf graviton}; $G$ $\rightarrow ~~(1_L,1_R)~+~2(
1_L,{1\over
2}_R)~+~(1_L,0_R)$

$\bullet$ {\bf gravitino}; $g_1$ $\rightarrow 2(1_L,{1\over
2}_R)~+~4(1_L,0_R)$

$\bullet$ {\bf gravitino}; $g_2$ $\rightarrow 2({1\over 2}_L,1_R)~+
{}~4({1\over 2}_L,{1\over 2}_R)~+~2({1\over 2}_L,0_R)$

$\bullet$ {\bf vector}; $~~~~V$  $\rightarrow ~~({1\over 2}_L,{1\over
2}_R)~+~2({1\over 2}_L,0_R)$

$\bullet$ {\bf tensor}; $~~~~T$ $\rightarrow ~~\,(0_L,1_R)~+~
2(0_L,{1\over 2}_R)~+~(0_L,0_R)$

$\bullet$ {\bf hyper}; $~~~~H$ $\rightarrow ~2(
0_L,{1\over 2}_R)~+~4(0_L,0_R)$
\vskip .5cm
All the above multiplets are obtained by tensoring Clifford algebra
states $(0_L,{1\over 2}_R)$, $2(0_L,0_R)$ by a $(J_L,J_R)$,
$SU(2)_L\times SU(2)_R$
representation, with a doubling (when required) by CPT.

The $N=(4,4)_G$ graviton multiplet decomposes into $N=(0,2)$ as
follows:
\vskip .5cm
$[N=(4,4)]_G~$ $=~(1_L,1_R) ~+~$ $ 4(1_L,{1\over 2}_R) ~+~$ $
4({1\over 2}_L,1_R)
{}~+~$ $ 16({1\over 2}_L,{1\over 2}_R) ~+~ $ $
 5(1_L,0_R) ~~~$

\hskip 3.cm +$5(0_L,1_R) ~+~$ $ 20({1\over 2}_L,0_R) ~+~$ $
20(0_L,{1\over 2}_R) ~+~$ $ 25(0_L,0_R)$.
\vskip .2cm
\hskip 2.5cm =  $1G~+~$ $1g_{1}~+~$ $2g_{2}~+~$ $8V~+~$
$5T~+~$ $5H$
\vskip .5cm
On the other hand we may decompose
$N=(4,4)\rightarrow N=(0,2)$ through the chain
$N=(4,4)\rightarrow N=(4,2) \rightarrow N(0,2)$.\\
We first get
$[N=(4,4)]_G \rightarrow [N=(4,2)]_G~+~[N=(4,2)]_g$,
($256\rightarrow 128 + 128$ states).\\
Then we can further decompose   $[N=(4,2)]_G$ and $[N=(4,2)]_g$ into
$N=(0,2)$ multiplets to finally obtain:
\vskip .3cm

$\bullet$ $[N=(4,2)]_G~=~(1_L,1_R)~+~$ $8({1\over 2}_L,{1\over
2}_R)~+~$ $(1_L,0_R)~+~$
$5(0_L,1_R)~+~$ $5(0_L,0_R)~~~$

\hskip 3.cm $~~~+4({1\over 2}_L,1_R)~+~$ $2(1_L,{1\over 2}_R)~+~$
$10(0_L,{1\over 2}_R)~+~$ $4({1\over 2}_L,0_R)$
\vskip.3cm
\hskip 3.cm =$[N=(0,2)]_G~+~2[N=(0,2)]_{g_2}~+~5[N=(0,2)]_T$

\vskip .3cm
$\bullet$ $[N=(4,2)]_g~=~2(1_L,{1\over 2}_R)~+~$ $8({1\over
2}_L,{1\over 2}_R)~+~$ $10(0_L,{1\over 2}_R)~~~$
$16({1\over 2}_L,0_R)~+~$

\hskip 3.cm +$4(1_L,0_R)~+~$ $20(0_L,0_R)$
\vskip .3cm
\hskip 3.cm $=~[N=(0,2)]_{g_1}~+~8[N=(0,2)]_V~+~5[N=(0,2)]_H$

\vskip.5cm
This implies, in  particular, the solvable Lie algebra decomposition
\be
{\rm \cal{S}}\left[{O(5,5)\over {O(5)\times O(5)}}\right]~=~
{\rm \cal{S}}\left[{O(1,5)\over { O(5)}}\right]~+~
{\rm \cal{S}}\left[{O(4,5)\over {O(4)\times O(5)}}\right]
\ee
i.e. algebras with a rank-1 tensor multiplet coset and a rank-4
quaternionic manifold with holonomy $SU(2)\times H$, $H=USp(2)\times
USp(4)$. This statement shows that the decomposition of the maximal
solvable Lie algebra into $N=2$ solvable Lie algebras corresponds to
the scalar sector of the two $N=(4,2)$ multiplets into which the
$[N=(4,4)]_G$ multiplet decomposes. If we delete the $[N=(4,2)]_{g}$
multiplet we obtain a pure $N=(4,2)$ supergravity theory with scalar
manifold ${O(1,5)/{ O(5)}}$.
This result was obtained by  Julia in \cite{jul}.

Note that a
similar procedure in $D=5$ and $D=4$ gives analogous relations among
solvable Lie algebras \cite{adaf}--\cite{gst}:
\be
{\rm \cal{S}}\left[{E_{6(6)}\over USp(8)}\right]~=~
{\rm \cal{S}}\left[{SU^*(6)\over USp(6)}\right]~+~
{\rm \cal{S}}\left[{F_{4(4)}\over {USp(2)\times USp(6)}}\right]
{}~~~~~~~ {\rm in}~D=5
\ee
\be
{\rm \cal{S}}\left[{E_{7(7)}\over SU(8)}\right]~=~
{\rm \cal{S}}\left[{SO^*(12)\over U(6)}\right]~~+~~
{\rm \cal{S}}\left[{E_{6(2)}\over {SU(2)\times SU(6)}}~\right]
{}~~~~~~~ {\rm in}~D=4.
\ee
The rank of the above cosets in the various dimensions decomposes as
follows:

$\bullet$ In $D=6$: $5~=~1V~+~4H$

$\bullet$ In $D=5$: $6~=~2V~+~4H$

$\bullet$ In $D=4$: $7~=~3V~+~4H$\\
This shows that the fields that correspond to the compactification
 radii
are vector multiplets; the rank of the hypermultiplet manifold does
not change with the dimensions. The rank-4 quaternionic spaces at
$D=6,5$ and 4, correspond to the following maximal decompositions of
the corresponding isometry groups:
\par
\par

$F_{4(4)}\rightarrow O(5,4), ~~~~~~~~ 52\rightarrow 36~+~16$.

$E_{6(2)}\rightarrow F_{4(4)}, ~~~~~~~~~~~ 78\rightarrow 52~+~26$.
\vskip .3cm

The total number of states of the $N=(4,2)$ supergravity in six
dimensions is 128 = 64 bosons + 64 fermions as in $D=4$ and $D=5$.
However the theory in $D=6$ is {\it anomalous} since the
gravitational anomaly is -360$I_{1\over2}$, where
$I_{1\over2}$ stands for the anomaly contribution of one Weyl fermion
in six dimensions. The gravitational anomaly contribution of the pure
$N=(4,0)$ supergravity multiplet is $-630I_{1\over 2}$, so that
$I_{(4,2)}={4\over 7}I_{(4,0)}$. In the case of the $(4,0)$ theory
the anomaly is cancelled by adding twenty-one $(4,0)$ tensor
multiplets, giving a contribution to the anomaly $+21\times 30
I_{1\over2}$ \cite{seib}. In the present case, owing to the absence
of
matter multiplets with $J_{max}$ less than ${3\over 2}$, the
$N=(4,2)$ cannot be anomaly-free. If one adds a $[N=(4,2)]_{g_1}$
multiplet then the supersymmetry is extended to $N=(4,4)$, which is
an
anomaly-free theory. It seems that this is the only consistent way to
cancel the anomaly. Another possibility to cancel the anomaly would
be to add twelve ~$(4,0)$ tensor multiplets; this solution, however,
explicitly violates $N=(4,2)$ supersymmetry.

It is of interest to understand the reason why the six- dimensional
theory with 24 supercharges is anomalous while, in lower dimensions,
theories with the same number of supercharges can be constructed in
type II string theories\cite{fk}.

In the next sections we will show the restoration of $N=(4,4)$ in six
dimensions, starting from a five-dimensional $N=6$ string theory and
decompactifying one of the dimensions considering either
$R_6\rightarrow \infty$ or $R_6\rightarrow 0$.

\section{$N=6$ construction in five dimensions}

In order to construct in string theory a ground state with
$N=(4_L+2_R)$
it is necessary to consider the type II theory with maximal
supersymmetry $N=(4_L+4_R)$; then one has to reduce the supersymmetry
asymmetrically by projecting out  half of the right moving
supercharges. The heterotic or type I construction of $N=(4_L+2_R)$
ground
states is impossible, since the maximal supersymmety is $N=(4_L+0_R)$
in
the heterotic theory and $N=4$ in that of type I.

We start by presenting the type II $N=(4_L+2_R)$ string models and
write
down their partition function. Models with the same number of
supercharges in $D=4$ have been obtained by applying projections to
the maximal-supersymmetry string theories \cite{fk}, which preserve
modular invariance as well as the conformal symmetries on the string
world-sheet. The maximal-supersymmetry type II model is described in
the light-cone gauge, by 8 world-sheet left/right-moving bosonic and
fermionic coordinates. In our notation, in $D=5$, the coordinates
$\psi_{\mu}^{L, R}$ and $X_{\mu}^{L, R}$ $(\mu = 3, 4, 5)$ represent
the space-time degrees of freedom (in the light-cone gauge), whereas
the remaining ones correspond to the fermionic $\chi_{I}^{L, R}$ and
bosonic $~\phi_{I}^{L, R},~I=1,2,3,4,5$ internal degrees of freedom.
This description will be appropriate for the asymmetric orbifold
construction\cite{nava}, where four of the right-moving internal
coordinates are twisted in order to reduce by a factor of 2 the
right-moving supersymmetries:
\be
Z_2~:~(\chi_R^I,~\phi_R^I)~\rightarrow~-(\chi_R^I,~\phi_R^I),
{}~~~~~I=2,3,4,5
\ee
In the fermionic construction\cite{abk},\cite{klt}, $\phi_I^{L, R}$
$(I=1,2,3,4,5)$ are replaced by a pair of Majorana--Weyl spinors
$y_I^{L, R}$ and $\omega_I^{L, R}$  $(I=1,2,3,4,5)$, by means of the
2d-boson-fermion equivalence:
\be
 J_I^{L,R}~=~\partial \phi_I^{L,R}~=~y_I^{L, R}~\omega_I^{L, R}.
\ee
The construction of string models amounts to a choice of boundary
conditions for the 2d fermions $\chi_{I}^{L, R},~ y_I^{L, R},~
\omega_I^{L, R}$, which satisfies local and global consistency
requirements\cite{abk}. The $N=(4_L+4_R)$ maximal supersymmetry
model,
constructed in this manner, have four space-time supercharges
originating from the left-moving sector and another four from the
right-moving sector. In the language of the fermionic
construction\cite{abk}, this model is defined by introducing three
basis sets, $F$, $S$ and $\bar S$. The first one contains all the
left- and the
right-moving fermions:
\be
F = [~\psi_{\mu}^L, \chi_I^L, y_I^L , \omega_I^L~|
{}~\psi_{\mu}^R, \chi_I^R, y_I^R, \omega_I^R~]
   ~~~~(\mu =3,4,5; I=1,..,5),      \label{fullset}
\ee
and the basis sets $S$ and $\bar{S}$, which contain only
eight left- or right-moving fermions and generate the
GSO projections of the maximal supersymmetry \cite{fk}:
\be
 S = [~\psi_{\mu}^L, \chi_I^L~] ~~~
\bar{S} = [~\psi_{\mu}^R, \chi_I^R~].  \label{es-es-bar}
\ee
Four of the gravitinos of the $N=(4_L+4_R)$ model belong to the
$S$-sector and the other four to the $\bar{S}$-sector. Then, by
applying the
appropriate projection one obtains the $N=(4_L+2_R)$ superstring
model.

A possible projection for constructing $N=(4_L+2_R)$ is specified by
a
choice of fermion basis $b$ such that
\be
 b = [~\psi_{\mu}^R, \chi_{1}^R, y_{2,3,4,5}^R~|~ y_1^L,
\omega_1^L,~ y_1^R, \omega_1^R ~].
                       \label{proj}
\ee
Then $b$ generates the desire left--right-asymmetric $Z_2$
projection.
Namely, it acts as a twist on $\chi_I^R$ and $J_I^R$ when
$I=2,3,4,5$. Also it acts non-trivially on the set of fermions
\be
T=~[ y_1^L, \omega_1^L,~ y_1^R, \omega_1^R ].
\ee
This action however keeps  the currents $J_1^{L,R}=\partial
\phi^{L,R}_1$ invariant. In the bosonic $\phi_1^{L,R}$ language, the
non-trivial action on $T$ implies a shifting by $1/2$ unit on the
$\phi_1$ lattice; $\phi_1^{L,R}~\rightarrow~\phi_1^{L,R} +\pi$.
Therefore, the $Z_2$ defined by $b$ acts as a shift on $\phi_1$ and
as an asymmetric twist on the remaining four internal
supercoordinates $\chi_I^R,~\phi_I^R,$ $I=2,3,4,5$. This asymmetric
$Z_2$ breaks half of the right-moving supersymmetries by projecting
out two of the gravitinos of the $\bar{S}$-sector. The resulting
model has the desired $N=(4_L+2_R)$ supersymmetry. The presence of
$T$
fermions in $b$ makes the states coming from the $b$-twisted sector
massive. Indeed, the lowest state in the $b$-twisted sector, has
right-moving conformal dimensions $\Delta_R~=~{10\over 16}$, which
corresponds to the conformal dimension of a spin field constructed
with the ten right-moving fermions $spin[~\psi_{\mu}^R, \chi_{1}^R,
y_{2,3,4,5}^R~ y_1^R, \omega_1^R ]$. Thus, the lower right-moving
mass level in this sector is $[m_R(b)]^2={10\over 16}~-{1\over
2}={1\over 8}$.

The absence of massless twisted sectors in {\it any consistent}
orbifold construction is expected. Indeed,  the the fact that
the massless
spectrum of $N=(4_L+2_R)$ in $D=5$ is fully determined by the
graviton
supermultiplet implies that all the $N=(4_L+2_R)$ massless states
must be the   $Z_2$-invariant states of  $N=(4_L+4_R)$ supergravity.
These states are all present in the untwisted sector and therefore
all  extra states have to be massive.

Using for instance,
\be
{\tilde b}= [~\psi_{\mu}^R, \chi_{1}^R, y_{2,3,4,5}^R~]
\ee
instead of $b$, then the $\tilde b$ asymmetric projection still
projects out the two right-moving gravitinos from the $\tilde
S$-sector. In this case, however, the $\tilde b$-twisted sector
contains
some extra massless states, since the conformal dimension of
$spin{[~\psi_{\mu}^R, \chi_{1}^R, y_{2,3,4,5}^R~]}$ is precisely 1/2.
Among the extra massless states of the $\tilde b$-twisted sector,
there exist
{\it two extra gravitinos} with vertex operator
\be
V^{\tilde b}_{3/2}=\Psi_{\mu}^{L}~spin{[~\psi_{\mu}^R, \chi_{1}^R,
y_{2,3,4,5}^R~]}_{+},
\ee
and so the $N=(4_L+2_R)$ supersymmetry is extended again to
the maximal $N=(4_L+4_R)$.

Already at this point one can guess the difficulties that will
arise in constructing a six-dimensional string model with
$N=(4_L+2_R)$
supersymmetry. Indeed, in six dimensions
the  coordinate  $\phi_1$ is non-compact and thus, the only available
asymmetric projection is the one that is based on $\tilde b$; this
projection, however, gives rise to  massless $\tilde b$ twisted
sectors,
which contain two extra right-moving gravitinos.

In the following, we will  discuss in more detail the $b$
fermionic model in $D=5$. Then this model will be generalized in
order to
include the compactification radius $R_1$ modulus associated to
$\phi_1$. Finally, in the next section, we study the extension of
supersymmetry in the two decompactification limits $R_1\rightarrow
\infty$ and $R_1\rightarrow 0$. We will show that the helicity
supertraces $B_{2n}={\rm str}~[s]^{2n}$ vanish for $n=0,1,2, 3$ in
both
decompactification limits, indicating the extension of
$N=(4_L+2_R)\rightarrow N=(4_L+4_R)$.

We start with the partition function of $N=(4_L+2_R)$  based on $b$:
$$
Z^{D=5}_{b} = {1\over {\rm Im}\tau^{3\over 2}
{}~\eta^{3}\bar{\eta}^{3}}~~
{1\over 4}~\sum_{a, b, {\bar a},{\bar b}= 0}^1
{(-)^{a + b + ab}\over \eta^4}~~
{(-)^{\bar{a}+\bar{b} +\bar{a}\bar{b}} \over \bar{\eta}^4}
$$
\be
{}~~~~~~~~~~~~~~~~~ ~~~~~~
\times {1\over 2}\sum_{h, g}~\theta[^a_b]^4 ~~  \bar{\theta}
[^{\bar{a}}_{\bar{b}}]^2
     ~\bar{\theta} [^{\bar{a}-h}_{\bar{b}-g}]
     ~\bar{\theta} [^{\bar{a}+h}_{\bar{b}+g}]~~Z_{5,5}[^h_g].
\ee
In the above expression $Z_{5,5}[^h_g]$ denotes the contribution of
the five compactified coordinates $\phi_I^{L,R},~I=1,...,5$
(fermionized); four of them are $(h,g)$-twisted while
$\phi_1^{L,R}$ is $(h,g)$-shifted. In terms of fermionic characters,
the analytic expression of $Z_{5,5}[^h_g]$ is:
\be
Z_{5,5}[^h_g]~=~ {1\over 2}\sum_{{\gamma}, {\delta}}
\left[{\theta[^{\gamma +h}_{\delta +g}]\bar{\theta}[^{\gamma
+h}_{\delta +g}] \over \eta \bar{\eta}} (-)^{\delta h+\gamma g
+hg}\right]\times
\left[{\theta[^{\gamma}_{\delta}]^4 \over \eta^4}
{}~{\bar{\theta}[^{\gamma}_{\delta}]^2 \bar{\theta}[^{\gamma
+h}_{\delta +g}]
 \bar{\theta}[^{\gamma -h}_{\delta -g}]
\over \bar{\eta}^4}(-)^{hg}\right]
\ee
where the term in the first square bracket corresponds to the
contribution of the compactified coordinate $\phi_1$ written in terms
of the fermionic characters of $y_1^{L,R}, \omega_1^{L,R}$. In terms
of $\phi_1$-lattice characters, it corresponds to a fixed $S^1$
radius at the fermionic point $R_1=1/\sqrt{2}$. The term in the
second square bracket corresponds to the characters of the $Z_2$
{\it asymmetric} orbifold, where four right-moving coordinates
$J_I\rightarrow -J_I, {}~I=2,3,4,5$ are twisted. The phase factors
$(-)^{\delta h+\gamma g +hg},~(-)^{hg}$ are dictated by modular
invariance.

As we have already explained, we would like to examine the behaviour
of the $N=(4_L+2_R)$ model in the decompactification limits
$R_1\rightarrow \infty$ and $R_1\rightarrow 0$. For this purpose we
must deform the above model and move away from the fermionic point
by switching on the marginal deformation that  is associated to the
modulus $R_1$.

Before doing that, let us first examine the moduli space of the above
model, namely all possible $J^L\cdot J^R~$ (1,1)-deformations which
we are able to switch on simultaneously. Due to the left--right
asymmetry the only possible deformations are:
$$
J_1^L\cdot J^R_1 \rightarrow ~R_1~~~~{\rm and} ~~~J_I^L\cdot
J^R_1\rightarrow~Y_I,~I=2,3,4,5.
$$
The first deformation corresponds to the $R_1$ modulus while
the remaining four correspond to the Wilson lines $Y_I$ moduli.
Altogether they form the perturbative moduli
space constructed from the  $NS$-scalars
\be
{\rm {\cal M}_p}~=~{O(5,1)\over O(5)}.
\ee
The perturbative moduli space is extended to $M_{np}={SU^*(6)/
USp(6)}$ once we include the dilaton moduli (singlet under $O(5)$)
and the remaining $RR$-scalars (in {\bf 15, 10}, and {\bf 10}'
representations of $O(5)$). The perturbative string states form the
charge lattice $\Gamma_{1,5}[^h_g]$ associated to the moduli space
${\cal M}_{p}$.

The absence of continuous deformations associated to the twisted
coordinates $\phi_I^R,~I=2,3,4,5$ implies that all radii
$R_I,~I=2,3,4,5$, of the ``twisted" four-dimensional space {\it must
be frozen} to some special values of the moduli space (the SO(8)
fermionic point in our case). This also means that the asymmetric
orbifold projection {\it is well defined only for special values of
the moduli} of the initially untwisted $\Gamma_{4,4}$ lattice, e.g.
the fermionic $SO(8)$ symmetric point.

Keeping this point in mind we can generalize the model to include
the radius and Wilson line deformations by replacing $Z_{5,5}[^h_g]$
by $Z_{5,5}(R_1, Y_I)$; since the non-vanishing Wilson lines do not
affect the decompactification limits we will restrict ourselves to
$Y_I=0$, keeping the modulus $R_1$ arbitrary:
\be
Z_{5,5}[^h_g](R_1, Y_I=0)~=~
{{\Gamma_{1,1}[^h_g](R_1) \over {\eta {\bar \eta}}}}
Z_{4,4}[^h_g].
\ee
where $Z_{4,4}[^h_g]$ is the asymmetric orbifold contribution
of the four internal coordinates $\phi_I,~~I=2,3,4,5$
\be
Z_{4,4}[^h_g]={1\over
2}\sum_{\gamma,\delta}\left[{\theta[^{\gamma}_{\delta}]^4 \over
\eta^4}
{}~{\bar{\theta}[^{\gamma}_{\delta}]^2 \bar{\theta}[^{\gamma
+h}_{\delta +g}]
 \bar{\theta}[^{\gamma -h}_{\delta -g}]
\over \bar{\eta}^4}(-)^{hg}\right]
\ee
and
$$
\Gamma_{1,1}[^h_g](R_1)=\sum_{m,n}
{\rm exp} \left[~i\pi gm~+~ i2\pi\tau P^2_L~
-i2\pi {\bar \tau} P_R^2\right]~~~~~~~~~~~
{}~~~~~~ ~~~~~~~~~~~~~~~
$$
with
\be
P_L={1\over 2}\left[~{m\over R_1}+{(2n+h)R_1\over 2}~\right],~~~~
P_R={1\over 2}\left[~{m\over R_1}-{(2n+h)R_1\over 2}~\right]
\ee
is the shifted $\phi_1$-lattice
\cite{koun},\cite{kirkoun},\cite{deco}.

The $N=(4_L+2_R)$ model constructed above is an {\it asymmetric
freely
acting orbifold}. Following refs. \cite{koun},
\cite{kirkoun},\cite{deco} it corresponds to a partial {\it
spontaneous} breaking \cite{koun},\cite{kirkoun} of supersymmetry
$N=(4_L+4_R)\rightarrow N=(4_L+2_R)$. The massless states of this
model
consist only of $N=(4_L+2_R)$ gravitational multiplet. The remaining
massless states of an $N=(4_L+4_R)$ (two 3/2-multiplets) become
massive,
with a common mass equal to:
$$
(m^2_{3/2})_{7,8} ~=~{1\over R_1^2}~~~~~~(h=0,~~|m|=1,~~n=0).
$$
These states correspond to the ``untwisted" sector states with
momentum $|m|=1$ and $2n+h=0$. The ``odd" winding states $h=1$
correspond to the twisted sector. The lowest twisted states
correspond to two massive 3/2-multiplets with masses proportional to:
$$
(m^2_{3/2})_{7',8'}~=~{R_1^2\over4}~~~~(h=1,~~m=0,~~|2n+h|=1)
$$

\section{Six-dimensional decompactification of the
five-dimensional $N=(4_L+2_R)$ string model}

A six-dimensional model in string theory can be defined from a
five-dimensional one in two different ways: either by sending the
compactification radius $R_1\rightarrow \infty$ or $R_1\rightarrow
0$. In the first case the Kaluza-Klein (KK) momenta become the
six-dimensional continuous momentum $P^2_6\sim m^2/R^2_1$. In the
limit
$R_1\rightarrow 0$, however, the KK states become superheavy while
the
string windings give rise to the six-dimensional continuous momentum
${\tilde P}^2_6 \sim n^2R^2$. Starting with the $D=5$, $N=(4_L+2_R)$
string defined in the previous section, we would like to examine the
six-dimensional theories obtained in the two stringy
decompactification limits.

The five-dimensional spectrum of this model  always contains $6$
massless gravitinos. There is also a tower of massive gravitinos with
the same $R$ symmetry charges having even KK-momentum $m$ and even
winding charge $n'=2n+h$ ($h=0$). When $R_1\rightarrow \infty$ or 0
one obtains $6$ massless gravitinos with continuous momenta, either
$P_6$ or $\tilde P_6$. However, there exist {\it two more extra
towers} of massive gravitinos with {\it different $R$-symmetry
charges}, which can become massless in the two decompactification
limits, namely:

$\bullet$ In the limit $R_1\rightarrow \infty$ the (two) massive
gravitinos with odd KK-momentum $m$ become massless,
$(m^2_{3/2})_{7,8}\rightarrow 0$, while the ``winding" gravitinos
$(m^2_{3/2})_{7',8'}\rightarrow \infty$ become superheavy and
decouple from the spectrum. The two extra massless gravitinos
$(m^2_{3/2})_{7,8}=0$ together with the six massless gravitinos of
$N=(4_L+2_R)$ restore in six dimensions the $N=(4_L+4_R)$
supersymmetry.

$\bullet$ In the other limit, $R_1\rightarrow 0$, the two
gravitinos with odd KK momentum $m$, $(m^2_{3/2})_{7,8}\rightarrow
\infty$ become superheavy and decouple while the gravitinos with
winding $n'=2n+h$ odd ($h=1$) become massless
$(m^2_{3/2})_{7',8'}\rightarrow 0$.

The ``field theory" analogue of the above model can be obtained by a
Scherk-Schwarz\cite{ssbreak} supersymmetry breaking mechanism
\cite{spbrsting},\cite{koun},\cite{kirkoun} starting either from
$N=2$ in $D=10$ or from $N=1$ in $D=11$. The difference from the
string model is the absence of the winding states,
$n'=2n+h=0$ ($n=0,~h=0$). Therefore, in field theory, the
first decompactification limit, $R_1\rightarrow \infty$, gives the
same
effective supergravity theory as in string theory. On the other hand
the second limit, $R_1\rightarrow 0$, does not correspond any more to
a
decompactification but rather to a ``dimensional reduction" where all
KK states are truncated out from the spectrum. In the limit
$R_1\rightarrow 0$ the field theory remains five-dimensional.

The restoration of the maximal supersymmetry in $D=6$ can be checked
in the two decompactification limits by constructing the helicity
supertrace $B_{2n}$
\cite{fsz},\cite{bkir},\cite{koun},\cite{kirkoun},\cite{deco} as a
function of the radius $R_1$:
\be
B_{2n}(R_1;t)={\rm tr}~(-)^{2s}~[s]^{2n}~e^{-\pi t
    M^2_s(R_1)}={\rm str}~[s]^{2n}~e^{-\pi t M^2_s(R_1)}.
\ee
The little group of massless particle in in D=5 is SO(3). By
``helicity" $s$ we mean $U(1)$ charge of a $U(1)$ sub-group of
$SO(3)$. $M_s(R_1)$ denotes the mass of the state.

If there is a restoration of $N=(4_L+4_R)$,
$B_6$ has to vanish in both decompactification
limits due to the $N=8$ supertrace identities:
$$
B_{2n}=0,~~~~~n=0,1,2,3.~~~~~{\rm in }~~~~ N=8.
$$
In $N=6$ $B_6$ is not trivial since the supertrace identities
for  $N=6$ are:
$$
B_{2n}=0,~~~~~n=0,1,2.~~~~~~{\rm in }~~~~~N=6.
$$
The supergravity massless sector of the $N=6$ supergravity gives a
non-trivial $B_6({\rm SUGRA})=45/2$. If in the decompactification
limits the $N=6$ supersymmetry is extented to $N=8$ due to the
presence of extra massless states, then  $B_6=0$  must be found
in both limits.

In order to derive $B_{2n}$ in string theory it is convenient to
define the helicity generating partition function
\cite{bkir},\cite{koun},\cite{kirkoun},\cite{R2corr},\cite{n3d4}:
\be
    Z^{\rm string} (v, \bar{v}) = {\rm Tr} ~q^{L_0}
{}~\bar{q}^{\bar{L_0}}
    ~e^{2\pi i v s_L - 2\pi i\bar{v} s_R },
{}~~q=e^{2\pi i\tau},~{\bar q}=e^{2\pi i\bar \tau},~~t={\rm Im}\tau,
                                    \label{generate}
\ee
where $s_L$ and $s_R$ denote the left- and right-moving helicities.
The physical helicity is given by $s= s_L + s_R$. Once $Z^{\rm
string} (v, \bar{v})$ is defined, then $B_{2n}$ can be derived from
$Z^{\rm string} (v, \bar{v})$ by the action of the left- and
right-helicity operators
\be
   Q = {1\over 2\pi i}{\partial \over \partial v},~~
   \bar{Q} = - {1\over 2\pi i}{\partial \over \partial \bar{v}},
                                 \label{partial}
{}~~
{\rm so~ that}~~
    B_{2n}^{\rm string}
           = ( Q + \bar{Q} )^{2 n}
           Z^{\rm string} ( v, \bar{v}) ~|_{v =\bar{v} = 0}.
\ee
The $v,\bar v$ helicity modifications for the partition function have
been studied earlier, in order to obtain exact solutions of string
theory in the background of magnetic fields and to
investigate the associated phase-transition
phenomena\cite{kounmag},\cite{kirkounreg},\cite{bkir}. In that
context the quantities $v$ and $\bar{v}$ play the role of the
background magnetic field.

An explicit expression for $Z^{\rm string} (v, \bar{v})$ for the
$N=(4,2)$ string model of the previous section is given by an
expression that is similar to the $v = \bar{v} = 0$ partition
function:

$$
Z^{\rm string} (v, \bar{v}) = {~\xi(v)~\bar{\xi}(\bar{v})\over
{}~\eta^{3}\bar{\eta}^{3}}~~
{1\over 4}~\sum_{a, b, {\bar a},{\bar b}= 0}^1
{(-)^{a + b + ab}\over \eta^4}~~
{(-)^{\bar{a}+\bar{b} +\bar{a}\bar{b}} \over \bar{\eta}^4}
{}~~~~~~~~~~~~~~~ ~~~~~~~
$$
\be
{}~~~~~~~~~~ ~~~~~
\times {1\over 2}\sum_{h, g}~
\theta[^a_b](v)
\bar{\theta}[^{\bar{a}}_{\bar{b}}](\bar v)
{}~\theta[^a_b]^3
\bar{\theta}[^{\bar{a}}_{\bar{b}}]
\bar{\theta} [^{\bar{a}-h}_{\bar{b}-g}]
\bar{\theta} [^{\bar{a}+h}_{\bar{b}+g}]~~
Z_{5,5}[^h_g](R_1),
\ee
where the $\xi (v)$ and $\bar{\xi}(\bar{v})$ modifications is due to
the helicity charge of the 2d bosonic oscillators. The $v, ~\bar v$
modifications due to the 2d fermionic degrees of freedom give rise to
non-zero characteristics to the fermionic
$\theta[^{\alpha}_{\beta}](v)$
and $\bar \theta [^{\bar \alpha}_{\bar \beta}](\bar v)$ functions
\cite{bkir},\cite{koun},\cite{kirkoun},\cite{R2corr},\cite{n3d4}. The
analytic expression of $\xi (v)$ is:
\be
\xi (v) = \prod_1^{\infty} { (1-q^n)^2 \over { (1-q^n e^{2\pi i v})
(1 - q^n e^{-2\pi i \bar{v}})}}
= {\sin \pi v\over \pi}
{\theta_1' \over \theta_1(v)},
\label{xi}
\ee
$~~~~~~~\xi (v) = \xi({-v}),~~\xi (0)=1.$

In the expression for $Z^{\rm string} (v, \bar{v})$
we can sum over the indices $(a, b)$ and $(\bar{a}, \bar{b})$
using  the Riemann identity of theta functions
\be
{1\over2}\sum_{a,b}
{(-)^{a + b + ab} \over \eta^4}
{}~\theta [^{a}_{ b}](v)
{}\theta [^{a}_{b}]
{}\theta [^{a}_{b}]
{}\theta [^{a}_{b}]
=~{1\over \eta^4}
 \theta[^1_1]\left({v\over 2}\right)
 \theta[^{1}_{1}] \left({v\over 2}\right)
 \theta[^{1}_{1}] \left({v\over 2}\right)
 \theta[^{1}_{1}] \left({v\over 2}\right),
\ee
and
$$
{1\over2}\sum_{\bar a,\bar b}
{(-)^{\bar a + \bar b +\bar  a \bar b} \over \bar \eta^4}
{}\bar \theta [^{\bar a}_{\bar  b}](\bar v)
{}\bar \theta [^{\bar a}_{\bar b}]~
{}\bar \theta [^{\bar a+h}_{\bar b+g}]
{}\bar \theta [^{\bar a-h}_{\bar b-g}]
= {1\over \bar \eta^4}
 \bar \theta[^1_1]\left({\bar v\over 2}\right)
\bar \theta[^{1}_{1}]
\left({\bar v\over 2}\right) ~
\bar \theta[^{1-h}_{1-g}] \left({\bar v\over 2}\right)
\bar \theta[^{1+h}_{1+g}] \left({\bar v\over 2}\right),
                 \label{leftpsi2}
$$
\be
{}~~~~
\ee
The vanishing of $B_0, B_1$ and $B_2$ follows automatically  from the
vanishing of $\theta[^1_1](v/2)$ and $\bar \theta[^1_1](\bar v/2)$
for $v=0$ and $\bar v=0$: Indeed, $Z^{\rm string} (v, \bar{v})$
vanishes like $v^4~\bar v^2$ in the sector with $(h,g)\ne(0,0)$
($N=6$ sector), and it vanishes like $v^4~\bar v^2$ in the sector
with $(h,g)=(0,0)$ ($N=8$ sector). Therefore, the only non-trivial
helicity supertrace is
\be
B_6~=~(Q+\bar Q)^6 Z^{\rm string} (v, \bar{v})|_{v=\bar v=0}~
=~15Q^4{\bar Q}^2 Z^{\rm string} (v, \bar{v})|_{v=\bar v=0},
\ee
receiving non-vanishing contribution from the $N=6$ sector with
$(h,g)\ne(0,0).$

Furthermore, using the identities
\be
4iQ~\theta[^1_1](v/2)~=~\theta[^0_0]~
\theta[^0_1]~\theta[^1_0]~=~2\eta^3,
\ee
the expression for $B_6$, as a function of
$R_1$\cite{n3d4},\cite{R2corr}, simplifies to:
\be
B_6= {45\over 4 }
 \sum_{(h,g)\neq (0,0)}~{\chi} [^h_g]
{}~\Gamma_{1,1}[^h_g]~|_{R_1},
\ee
where $\Gamma_{1,1}[^h_g]~|_{R_1}$ is the radius-dependent shifted
lattice and ${\chi} [^h_g]$ \cite{n3d4},\cite{R2corr} is a
holomorphic
function coming from the four left-moving invariant coordinates
$\phi^L_I,~I=2,3,4,5$:
\be
{\chi} [^h_g]~=~{1\over 2}
\sum_{\gamma,\delta}~\theta[^{\gamma}_{\delta}]^4
\left[e^{i\pi(h+g\gamma)}~-
{}~e^{i\pi(g+h\delta)}\right].
\ee
In the infrared limit Im$~\tau \rightarrow\infty$ only the massless
states give a non-zero contribution for any finite $R_1$. In this
limit
\be
B_6(t \rightarrow\infty)~=45/2=-360~s_{1/2},
{}~~~~~~t\equiv{\rm Im}~\tau
\ee
$ s_{1/2}=-1/16$ denotes the contribution to str$[s]^6$ of a massless
spin-1/2 state. Thus, $B_6(t \rightarrow\infty)$ matches the
contribution of the massless fields of the $N=6$ supergravity
multiplet $B_6^{N=6}( {\rm SUGRA})=-360~s_{1/2}$. Therefore for any
finite $R_1$ the massless degrees of freedom are precisely those of
the $N=(4_L+2_R)$ supergravity multiplet in $D=5$.

Having obtained $B_6$ as a function of $R_1$, we can examine the
limits
$R_1\rightarrow \infty$ and $R_1\rightarrow 0$. In the first limit
the winding states are superheavy and give exponentially suppressed
contributions to $B_6$. Only the zero winding sector survives:
\be
B_6(R_1\rightarrow \infty)={45 \over4}~{\chi[^0_1]}~
\Gamma[^0_1]=
{45 \over 4}~\left[\theta^4_3(\tau)+
\theta^4_4(\tau)\right]~\theta_4\left
[{it \over R^2_1}\right]=
{45\over 2}~\theta_4\left
[{it\over R^2_1}\right]
\ee
where the last equality follows from the left/right mass-level
matching condition in the zero winding sector. Performing a Poisson
resummation of the last expression we find
\be
(B_6)_{D=5}~=~ R~t^{-{1\over 2}} ~(B_6)_{D=6},
\ee
with
\be
(B_6)_{D=6}={45\over 2}~\theta_2\left[{iR_1^2\over t }\right] ~
\sim~{45\over 2}~{\rm \cal O}\left (~2{\rm exp}{-\pi
R^2_1\over 4t}\right ).
\ee
In the above equation we renormalize the six-dimensional
$(B_6)_{D=6}$ by dividing out the $S^1$ volume $R_1$ and the
contribution of the zero modes of a non-compact coordinate $
t^{-{1\over 2}}$. Thus we show that $B_6$ vanishes exponentially in
the decompactification limit $R_1\rightarrow \infty$ due to the
restoration of $N=8$ in $D=6$.

 When $R_1\rightarrow 0$ the KK momenta become superheavy, so
that only
the sector with zero  KK-momenta ($m=0$) survives:
$$
B_6(R_1\rightarrow 0)=
{45\over 4}~\sum_{(h,g) \ne (0,0)}\chi[^h_g]
{}~\Gamma[^h_g]|_{m=0}
{}~~~~~~~~~~~~~~~~~~~~~~~~~~~~~~~~~~~~~~~~
$$
$$
={45\over 4}~\left(\chi[^1_0]+\chi[^1_1]\right)~
\theta_2 \left [i t R^2_1\right]+
{45\over 4}\chi[^0_1]~\theta_3\left[i t R^2_1\right]
{}~~~~~~~~~~~~~~~~~~~~~~~
$$
\be
{}~~~~={45\over 2}\left (~\theta_3 \left[it R^2_1\right]-
\theta_2 \left[ it R^2_1\right]~\right),
\ee
where we have used the identity $\chi[^0_1]+\chi[^1_0]+\chi[^1_1]=0$,
and the left/right-mass level matching conditions are  valid in the
$m=0$
sector. Performing a Poisson resummation of the above expression for
$B_6$ we obtain:
\be
(B_6)_{D=5}={t^{-{1\over 2}}\over R_1}(B_6)_{D=6}
\ee
with
\be
(B_6)_{D=6}~=~{45\over 2}~\left (~\theta_3 \left[{i\over
tR^2_1}\right]-
\theta_4 \left[ {i\over tR^2_1}\right]~\right)~\sim ~{45\over 2}~{\rm
\cal O}
\left (~2{\rm exp}~{-\pi\over ~tR_1^2}~\right ),
\ee
where the normalization factor from $D=5$ to $D=6$  now  involves the
dual $S^1$ volume $1/R_1$. As in the previous decompactification
limit, $B_6(R_1\rightarrow 0)$ vanishes exponentially in the dual
decompactification limit $(R_1\rightarrow 0)$ due to the restoration
of $N=8$ in $D=6$.

\section{ Conclusions}
\par
In this paper we have given evidence for the existence of a chiral
$N=(4,2)$ supergravity in six dimensions, which reduces to known
$N=6$ supergravity upon dimensional reduction at $D=5$ and $D=4$.

This theory is based on an automorphism group of the supersymmetry
algebra, which is $H=USp(4)\times USp(2)$ and a ``duality group'',
which is $O(1,5)$. Therefore the fermions of this theory are in
$H$-representations, the vectors are in the $O(1,5)$ spinor
representation, the two-forms  are in the vector representation
of $O(1,5)$ and the scalars are coordinates of the
$O(1,5)/O(5)$ coset. This is a consequence of the $(0,2)$
decomposition of the $(4,2)$ supergravity multiplet, which reveals
that the scalars are in five tensor multiplets \cite{rom}.

This theory, like $(2,0)$ and $(4,0)$ $D=6$ supergravities, is a
non-Lagrangian theory, due to the presence of self-dual $2$-forms,
but it can presumably be defined through a consistent set of
equations of motion, by using supersymmetry. However, unlike the
other chiral
theories, this theory has a gravitational anomaly that cannot be
cancelled by adding matter, since there are no $N=(4,2)$ multiplets
with spin $J<3/2$.

As a consequence the $N=(4,2)$ theory is inconsistent at the  quantum
level. A way  the
anomaly cancellation can be achieved is, either by adding another
gravitino
multiplet, which enlarges the theory to $N=(4,4)$, or by adding
twelve
$N=(4,0)$ matter multiplets, which are expected to explicitly break
two supersymmetries and then make the theory inconsistent.

It is interesting to understand why the analogous theories at $D=4$
and $D=5$ are consistent string backgrounds while the $D=6$ case is
not. If we define the $D=6$ theory by decompactifying one dimension
from the $D=5$ case, then some extra massive states become massless
in the decompactification limits $R_6\rightarrow \infty$ or 0. In
both limits the extra massless states correspond to an extra
$N=(4,2)$ gravitino multiplet, which cancels precisely the
six-dimensional anomaly and the $B_6$-helicity supertrace at the same
time. The supersymmetry is extended to the maximal one and the theory
becomes left/right-symmetric.

\vspace*{.4cm}

\noindent
{\bf Acknowledgements}
\vspace*{.3cm}

We would like to thank L. Andrianopoli, M. Flato and
C. Fronsdal for interesting discussions.

This work is supported in part by EEC under TMR contracts
ERB-4061-PL-95-0789, ERBFMRX-CT96-0045 and by
DOE grant DE-GG03-91ER40662.

\vskip .4cm
\vfil
\eject


\end{document}